# Mesoscopic Coulomb Blockade in One-channel Quantum Dots

S. M. Cronenwett, S. M. Maurer, S. R. Patel, C. M. Marcus
*Department of Physics, Stanford University, Stanford, California 94305*

C. I. Duruöz, J. S. Harris, Jr.
*Department of Electrical Engineering, Stanford University, Stanford, California 94305*

Signatures of "mesoscopic Coulomb blockade" are reported for quantum dots with one fully transmitting point-contact lead, $\mathcal{T}_1 = 1$, $\mathcal{T}_2 \ll 1$. Unlike Coulomb blockade (CB) in weak-tunneling devices ($\mathcal{T}_1$, $\mathcal{T}_2 \ll 1$), one-channel CB is a mesoscopic effect requiring quantum coherence. Several distinctive features of mesoscopic CB are observed, including a reduction in CB upon breaking time-reversal symmetry with a magnetic field, relatively large fluctuations of peak position as a function of magnetic field, and strong temperature dependence on the scale of the quantum level spacing.

Quantum dots provide a simple system to probe both quantum mechanical interference effects and the physics of electron-electron interactions. In dots weakly coupled to electronic reservoirs, i.e. when the two leads have transmissions $\mathcal{T}_1$, $\mathcal{T}_2 \ll 1$, transport measurements show the influence of electron-electron interactions in the form of Coulomb blockade (CB), a classical effect which appears as strong peaks in dot conductance as an external gate voltage is swept, changing the dot potential [1]. Quantum interference in the weak-tunneling regime leads to random fluctuations of the CB peak height as a function of magnetic field and dot potential [2, 3]. In the opposite limit of several open channels per lead, $\mathcal{T}_1$, $\mathcal{T}_2 > 1$, quantum levels broadly overlap and a semiclassical picture of interfering trajectories of charged particles scattering through the dot can be used to describe transport. In the open regime, the physics of quantum interference is well understood, leading to coherent backscattering at zero magnetic field and universal conductance fluctuations [4]. Electron-electron interactions do not strongly affect transport in open dots, and no CB is observed. In the crossover regime, where the transmission of at least one point contact is unity, CB in dots disappears classically [5, 6] but can appear as a result of quantum interference, as recently discussed in Ref. [7]. Such "one-channel CB" will be the subject of this Letter.

In metallic dots, it is known both theoretically [8] and experimentally [9] that CB oscillations are washed out as the tunneling strength increases above $\mathcal{T} = 1$ in either lead. In semiconductor quantum dots, experiments of CB in the strong tunneling regime have yielded varying results. Kouwenhoven, *et al.* [10], found that CB disappears when the transmission through one point contact reaches unity. However, Pasquier, *et al.* [11] report small CB oscillations up to $\mathcal{T}_1 + \mathcal{T}_2 \sim 3$, and Crouch, *et al.* [12] observed CB which decreases near $\mathcal{T}_1 = 1$, $\mathcal{T}_2 \ll 1$, but then increased again above $\mathcal{T}_1 > 1$. Flensberg [5] and Matveev [6] have shown theoretically that CB disappears at unity transmission when only inelastic processes are taken into account, appropriate in the limit $kT/\Delta \gg 1$, where $\Delta$ is the quantum level spacing. Recently, Aleiner and Glazman [7] extended the analysis to include elastic processes, finding for the particular case of "one-channel" transport ($\mathcal{T}_1 = 1$, $\mathcal{T}_2 \ll 1$) that although open-channel CB vanishes for purely inelastic transport, it persists due to coherent mechanisms for temperatures comparable to the quantum level spacing.

In this Letter, we investigate several novel features of one-channel CB which illustrate the interplay between quantum interference and electron-electron interactions. These are: (1) an enhancement of CB around zero magnetic field which, like coherent backscattering in open structures, can be understood in terms of the breaking of time-reversal symmetry; (2) a strong temperature dependence of CB on the scale of the quantum level spacing; (3) an enhanced correlation (though less than predicted theoretically) of conductance as a function of gate voltage, compared to the weak-tunneling regime, which results from cotunneling through many levels both on and off the CB peak; (4) large peak motion, predicted to be on the scale of the separation of CB peaks, as a function of magnetic field.

The existence of CB in the one-channel regime can be understood in terms of an effective scatterer at the location of the open lead that arises due to coherent trajectories reflected from the walls of the dot [7]. It is known that a real scatterer in a nearly-open lead ($T_1$ below unity) will cause classical CB [5, 6]. Coherent processes give rise to a standing wave at the open lead which acts as such a scatterer, also giving rise to CB [7]. We emphasize that without coherence, CB should disappear in the fully transmitting one-channel case. At zero magnetic field, backscattering increases due to constructive interference of time-reversed paths, leading to stronger CB oscillations. As the magnetic field $B$ is increased beyond a characteristic field $B_C$ (where $B_C$ puts of order one flux quantum through the backscattered trajectories) the spectral power in the CB oscillations, $P_{CB}$, is predicted to decrease by a factor of 4 [7]. The requirement of coherence for CB in the one-channel dot implies a strong temperature dependence of $P_{CB}$ on the scale of the quantum level spacing $\Delta$ rather than the classical charging energy. Detailed calculations yield

$$P_{CB}(B,T) = \frac{G_R^{\,2}}{4} \Lambda(B)\, \alpha\, \frac{\Delta}{T}\left(\frac{\Delta}{E_C}\right)^2 \ln^3\left(\frac{E_c}{2\pi kT}\right), \tag{1}$$

where $\Lambda(B \ll B_C) = 4$ and $\Lambda(B \gg B_C) = 1$, $E_C$ is the classical charging energy, i.e. the energy to add a single electron to the dot, $\alpha \approx 0.207$ is a numerical factor, and $G_R$ is the conductance of the tunneling point contact [7]. Another consequence of the coherent nature of one-channel CB is that small changes in parameters such as device shape or magnetic field, which alter the interference pattern in the dot, can shift the position in gate voltage where the CB peak appears.

Experimentally this shows up as a strongly *B*-dependent peak position, with excursions on the scale of the spacing between peaks.

We report measurements of two quantum dots [micrograph in Fig. 3(d), inset] fabricated using CrAu electrostatic gates 900 Å above a two-dimensional electron gas (2DEG) on a GaAs/AlGaAs heterostructure. A multiple-gate design allows independent control of point contact conductances and dot shape via several shape-distorting gates. Both dots have an area of 0.5 µm² giving a level spacing $\Delta = 2\pi\hbar^2/m^* A$ = 14 µeV (m* is the effective electron mass, *A* is the dot area assuming a 100 nm depletion width). Measurements were made in a dilution refrigerator with an ac voltage bias of ~ 5 µV at 13.5 Hz. The experimental temperature *T* used throughout refers to the electron temperature, measured from the widths of CB peaks. At fridge base, *T* = 100 mK. Gate voltages can be related to dot energy through the ratio $\eta$, measured from the linear temperature dependence of the full-width-at-half-max of CB peaks in the weak-tunneling regime at $T > \Delta$, where $e\eta$(FWHM) ≈ $4.3 k_B T$. This ratio then converts peak spacing to charging energy, giving $e\eta$(*peak spacing*) = $E_C$ = 260 µeV for dot 1 and $E_C$ = 320 µeV for dot 2.

Figure 1 illustrates a number of novel features of one-channel transport. The rapid oscillations in conductance as a function of gate voltage are the CB oscillations. Comparing Figs. 1(a) and 1(b) shows that CB oscillations in the one-channel regime are considerably stronger at *B* = 0 mT ( << $B_C$) compared to 100 mT ( >> $B_C$ ~ 20 mT), unlike the weak-tunneling regime [Figs. 1(c, d)], where the strength of CB does not appear to depend on magnetic field. One-channel CB also shows large fluctuations of valley conductance due to large cotunneling contributions which are suppressed in the weak-tunneling regime.

The *B* dependence of one-channel CB arising from the breaking of time-reversal symmetry can be studied quantitatively by evaluating the power spectral density $P_g(f)$ of the conductance, $g(V_g)$, at a number of different magnetic fields. The argument *f* is the gate-voltage frequency in units of cycles/mV [13]. In both the one-channel and weak-tunneling regimes, $P_g(f)$ shows a clear peak around the CB frequency, $f_{CB} = \eta/E_C$ , as seen in Figs. 2(a, b). In the one-channel regime, the CB peak in $P_g(f)$ has a clear maximum around $B = 0$, whereas CB in the weak-tunneling regime is essentially independent of $B$. We define the CB power, $P_{CB}$, as the power in a window around the CB frequency in $P_g(f)$ (bracketed regions in Figs. 2(a, b)). The enhanced CB power around $B = 0$ in the one-channel regime now appears as a peak at $B = 0$ in the function $P_{CB}(B)$ [Fig. 2(c)], while $P_{CB}(B)$ is flat for weak tunneling [Fig. 2(d)]. We note that the width of the CB peak in $P_g(f)$ around $f_{CB}$ is broader and shows greater fluctuations in frequency in one-channel CB compared to the weak-tunneling regime. This implies a broader distribution of peak spacings in the one-channel regime, contrary to the predictions of Ref. [7], and remains an interesting open problem.

CB power normalized by its large-$B$ average, $p_{CB}(B) = P_{CB}(B)/\langle P_{CB}(B)\rangle_{B>>B_c}$, and averaged over an ensemble of several dot shapes provides a useful quantity for comparing the $B$ dependence of CB in the one-channel and weak-tunneling regimes [Fig. 3(a)]. To compute these data, $P_{CB}(B)$ traces from seven (three) data sets in the one-channel (weak-tunneling) regime were normalized by the average value over 27 mT < $B$ < 130 mT [for two of the one-channel sets: 21 mT < $B$ < 60 mT] and then averaged. The zero-field value, $\langle p_{CB}(0)\rangle \sim 5.3 \pm 0.5$, is somewhat larger than the predicted factor of 4 for reasons not yet understood. In the weak-tunneling regime $\langle p_{CB}(0)\rangle \sim 0.7 \pm 0.2$, somewhat closer to unity than the zero-temperature theoretical value of 9/16 [2], presumably due to decoherence. We note that a $B$-dependent real reflection in the point contact with maximum reflection at $B = 0$ could lead to a spurious enhancement of CB power at $B = 0$. To rule out this possibility, we have measured (in a separate device) the field dependence of the open point contact with the rest of the dot undepleted and find only very slight $B$ dependence with no distinct features on the 10-40 mT scale.

The temperature dependence of one-channel CB power is shown in Fig. 3(b) along with the no-free-parameters theory, Eq. (1), for both $B << B_C$ and $B >> B_C$. Experiment and theory are roughly consistent up to $T \sim 300$ mK ($kT \sim 2.5\, \Delta$), with good agreement in slope and the $T$ independence of the ratio $P_{CB}(B << B_C)/P_{CB}(B >> B_C)$, and reasonable agreement in absolute magnitude given the lack of free parameters. Note the log scale on the vertical axes and that CB powers range over a factor of ~ 100. At temperatures above ~ 400 mK, one-channel CB is strongly suppressed and the enhancement around $B = 0$ disappears, as seen in both Figs. 2(b) and 2(c).

We have also investigated correlations in conductance as a function of $V_g$, $C(\delta V_g) = \langle g(V_g)g(V_g + \delta V_g)\rangle_{V_g}$ (average is over gate voltage) in the one-channel regime. Long correlations are expected in one-channel CB due to the significant contributions of cotunneling via many levels (of order $E_C/\Delta$) [14]. Figure 3(d) shows the discrete correlation function $C(\delta n)$, where $\delta n = 0,1,2,...$ acts as a peak index, defined for the one-channel CB regime as $C(\delta V_g)$ evaluated at the CB period, $\delta V_g = \delta n(E_C/\eta)$. In the weak-tunneling regime, $C(\delta n)$ is directly evaluated using sets of discrete peaks heights, $C(\delta n) = \langle g_{max}(n)g_{max}(n+\delta n)\rangle_n$, to avoid spurious correlations in $C(\delta V_g)$ caused by uniformly low valley conductance. The correlation length in the one-channel CB is considerably shorter than the theoretical value of $E_C/\Delta \sim 15$ peaks, and not significantly different than that of the weak-tunneling regime. This short correlation may be caused by changes in the energy spectrum of the dot as electrons are added with each successive CB peak. This suggests that the number of added electrons sufficient to scramble the dot spectrum is less than $E_C/\Delta$, consistent with similar conclusions based on the temperature dependence of peak correlations in the weak-tunneling regime [15].

Finally, we have investigated the predicted large-scale peak motion as a function of magnetic field in the one-channel CB regime. Whereas the weak-tunneling regime exhibits CB peak motion on the scale of the level spacing (once scaled to dot energy by $\eta$) [3], peak motion in the one-channel regime is expected to be of order $E_C$ [7]. An enhanced peak motion in $B$ for one-channel CB compared to weak-tunneling CB is seen in Fig. 4, although the effect is not as large as predicted theoretically. The standard deviation of peak motion about its average position is $0.09 E_C$ ( ~ 2 $\Delta$) for one-channel CB [Fig. 4(a)], compared to $0.02 E_C$ (~ 0.5 $\Delta$) in the weak-tunneling regime [Fig. 4(b)], the latter consistent with previous measurements [3]. Large-scale motion of CB peaks in the strong-tunneling regime is the subject of ongoing investigations.

In summary, we have presented measurements of mesoscopic Coulomb blockade which arises due to quantum coherence in a quantum dot with one fully transmitting channel ($\mathcal{T}_1 = 1$, $\mathcal{T}_2 \ll 1$). One-channel CB is enhanced in the presence time-reversal symmetry and has a strong temperature dependence on the scale of the quantum level spacing, consistent with theory [7]. Correlation of conductance in gate voltage appears limited to ~ 3 peaks, smaller than expected, perhaps as a result of changes in the energy spectrum of the dot upon adding electrons. The motion of mesoscopic CB peaks with magnetic field is significantly greater than in the weak-tunneling regime, but smaller than expected theoretically.

We thank I. Aleiner, L. Glazman, A. Johnson, L. Kouwenhoven, and K. Matveev for valuable discussions. We acknowledge support from the ARO under DAAH04-95-1-0331, the ORN-YIP under N00014-94-1-0622, the NSF NYI and PECASE programs under DMR 9629180-1, the A. P. Sloan Foundation (Marcus Group), and JSEP under DAAH04-94-G-0058 (Harris Group). SMC was supported by an NSF Graduate Fellowship and SMM was supported by a Hertz Foundation Fellowship.

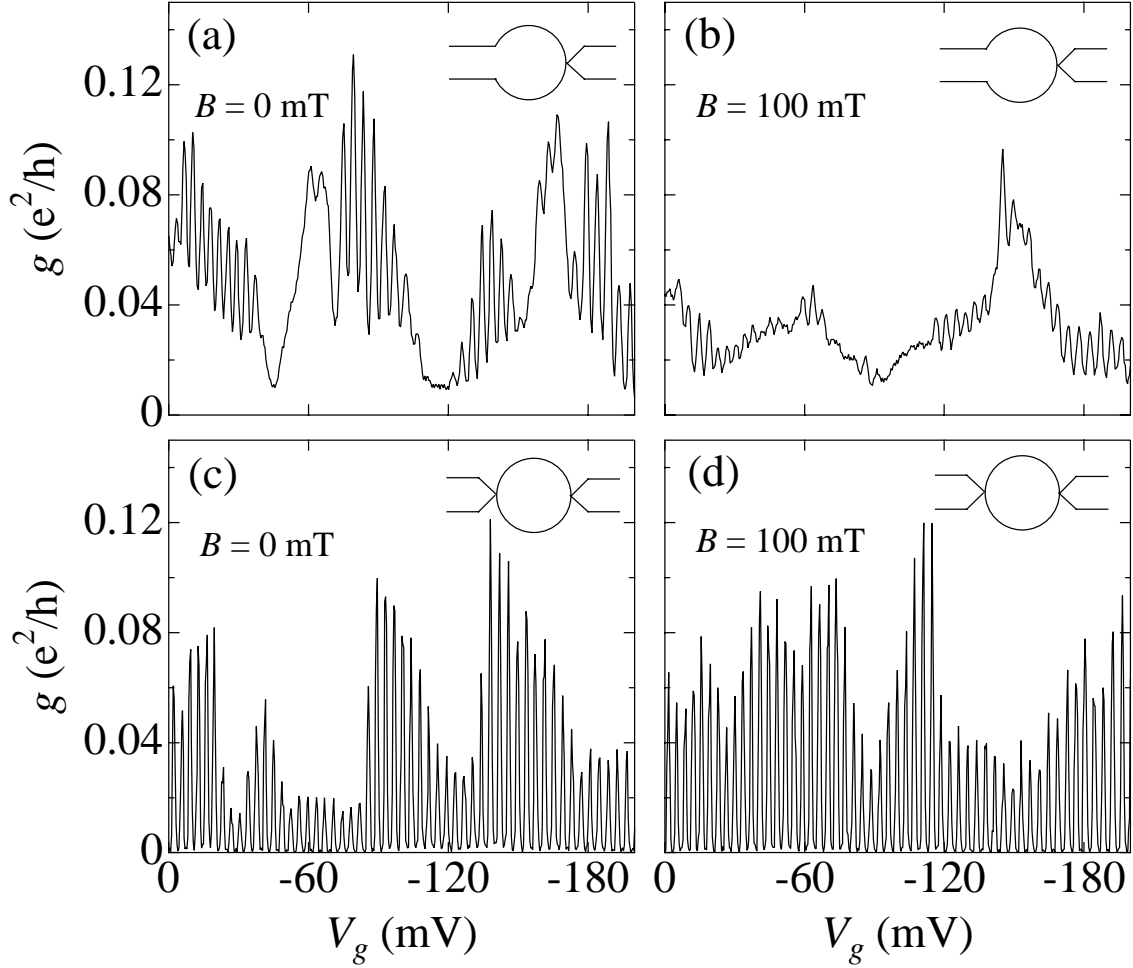

**Fig. 1**: Conductance showing Coulomb blockade (CB) oscillations as a function of gate voltage, $V_g$, in the one-channel regime (a, b) and weak-tunneling regime (c, d) at $B = 0$ mT ($< B_C \sim 20$ mT) and $B = 100$ mT ($> B_C$) (dot 1). One-channel CB is stronger at $B = 0$ mT compared to $B = 100$ mT, unlike weak-tunneling CB.

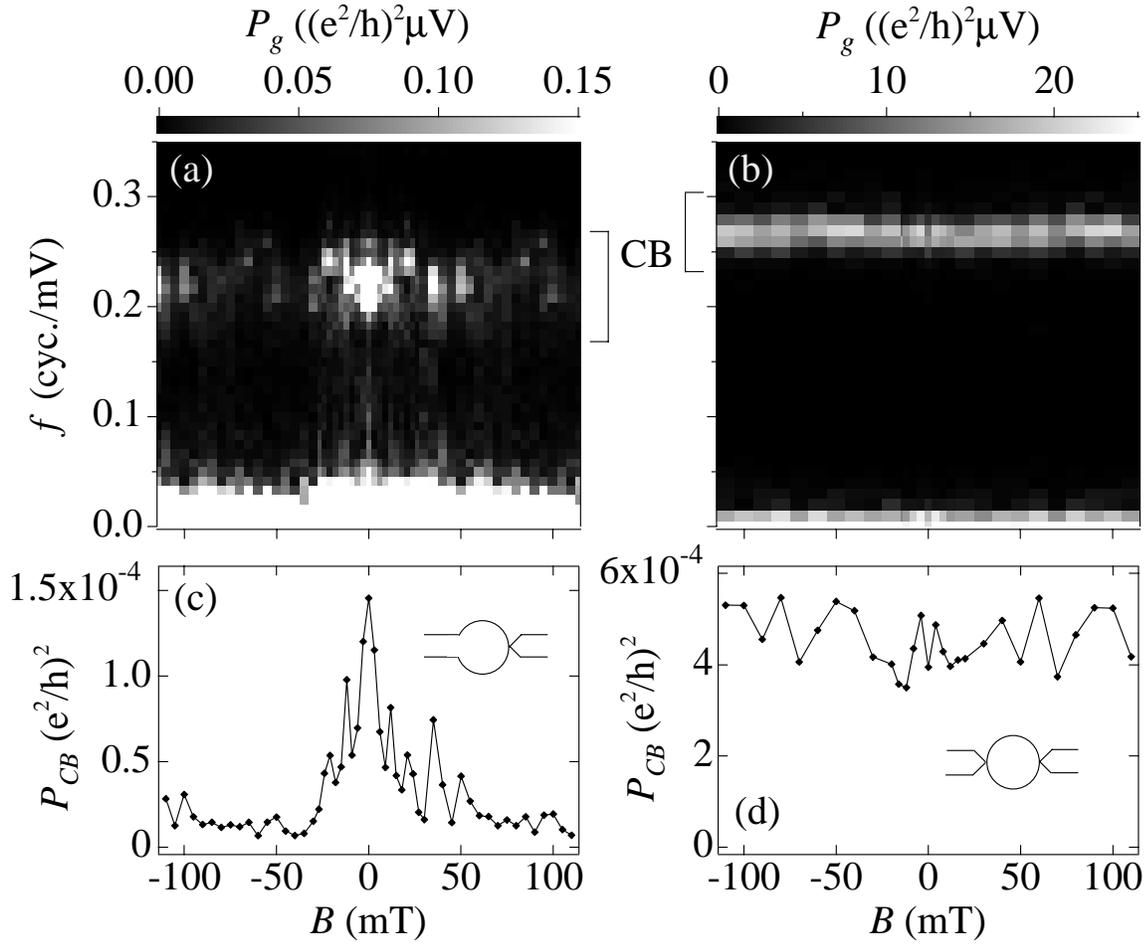

**Fig. 2**: (a, b) Grayscale plots of the power spectral density, $P_g$, as a function of gate-voltage frequency $f$ (cycles/mV) and magnetic field $B$ for (a) one-channel CB and (b) weak-tunneling CB (dot 1). The dominate frequency of CB oscillations is within the bracketed region marked "CB" in each plot. In the one-channel regime (a), the bright structure at $B \sim 0$ at the CB frequency indicates stronger CB at zero field. No corresponding field dependence of CB is seen for weak tunneling (b). At each magnetic field, the power within the bracketed region defines $P_{CB}$ $(B)$, the CB power. (c, d) $P_{CB}(B)$ for data in (a) and (b). Again, the zero-field enhancement of CB in the one-channel regime is seen as a peak in $P_{CB}(B)$ around $B = 0$.

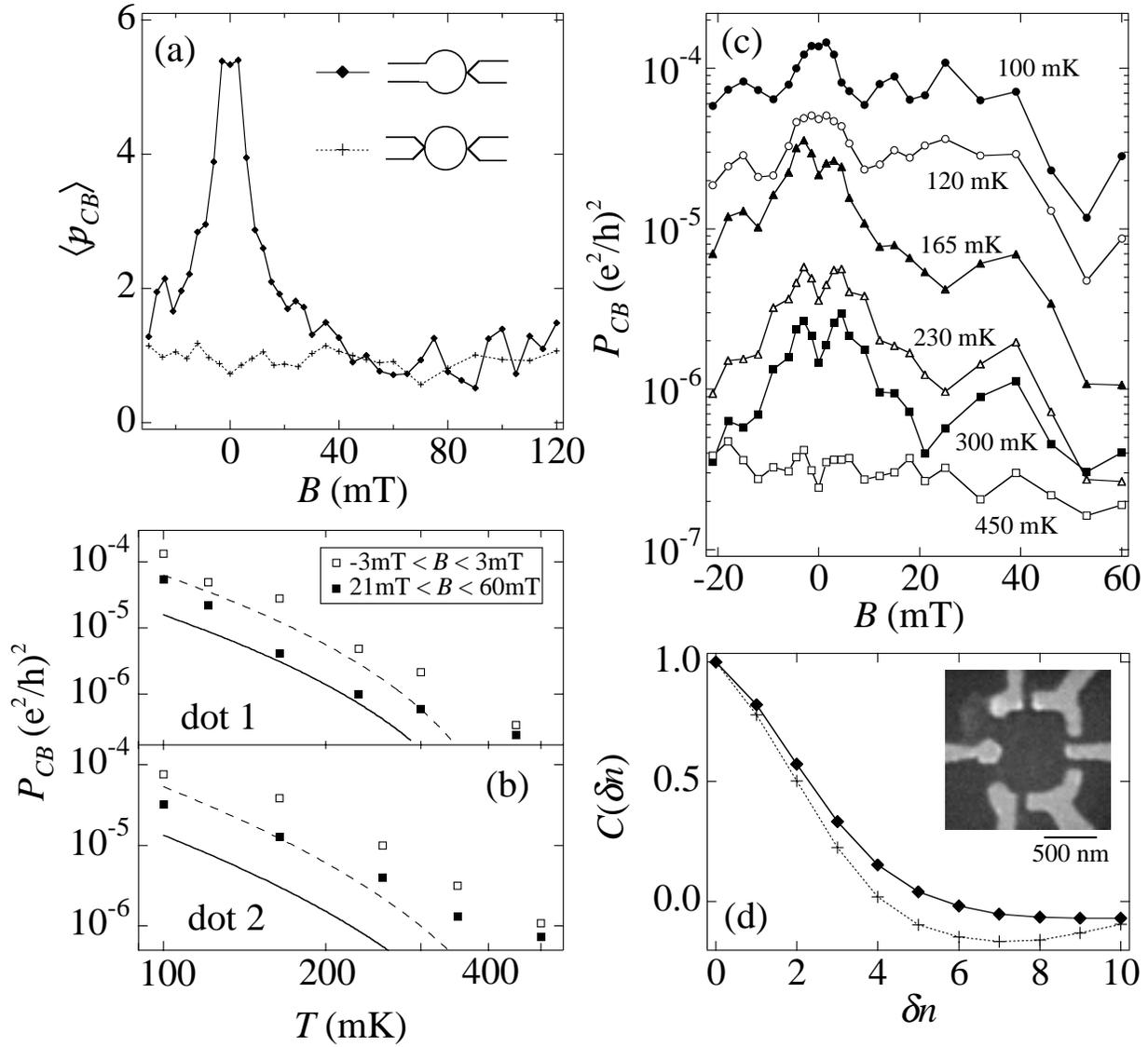

**Fig. 3:** (a) Magnetic field dependence of CB oscillations for the one-channel and weak-tunneling regimes given by the CB power normalized to its $B \gg B_C$ value and averaged over shape configurations, $\langle p_{CB}(B) \rangle$ (dot 1). Uncertainties are ~10% for the one-channel regime and ~30% for the weak-tunneling regime. (b) CB power, $P_{CB}(B)$, as a function of temperature, averaged over field ranges $-3$ mT $< B < 3$ mT (open squares) and $21$ mT $< B < 60$ mT (solid squares), along with theory (Eq. (1)) for $B \ll B_C$ (dashed curve) and $B \gg B_C$ (solid curve). (c) $P_{CB}(B)$ decreases strongly with increasing temperature from 100 mK (top curve) to 450 mK (bottom curve) in dot 1. The zero-field peak in $P_{CB}(B)$ persists up to ~ 400 mK. (d) Magnetic-field-averaged autocorrelation function, $C(\delta n)$, of CB conductance oscillations in gate voltage $V_g$ in units of peak number $\delta n$, where $\delta V_g = \delta n (E_C/\eta)$ for dot 1. The correlation length for the one-channel CB regime (diamonds) is ~ 3 peaks, slightly larger, but comparable to the weak-tunneling case (crosses). Inset: SEM micrograph of dot 1.

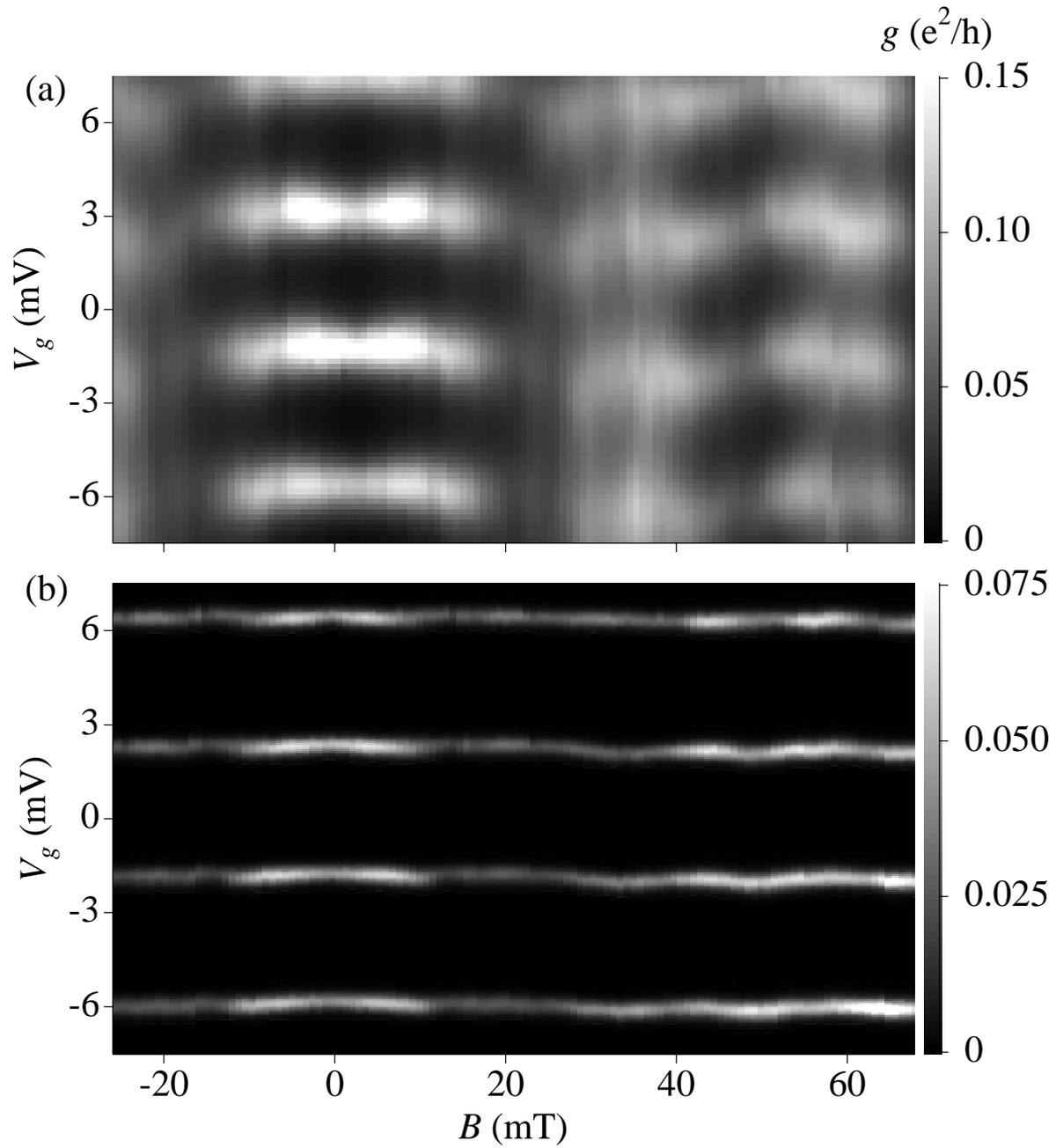

**Fig. 4**: Grayscale plots of conductance versus gate voltage $V_g$ and magnetic field $B$ shows larger fluctuations of CB peak position as a function of $B$ in the one-channel regime (a) than in the weak-tunneling regime (b). Note symmetry in $B$ of CB peak height and position (dot 1).